# Advances in Crystallographic Image Processing for Scanning Probe Microscopy

**P. Moeck**

Nano-Crystallography Group, Department of Physics, Portland State University, 1719 SW 10$^{th}$ Avenue (SRTC), Portland, OR 97201, USA, pmoeck@pdx.edu

This book chapter reviews progress in crystallographic image processing (CIP) for scanning probe microscopy (SPM) that has occurred since our description of the technique was first put into open access in this book series in the year 2010. The signal to noise ratio in all kinds of experimental images of more or less regular 2D periodic arrays is significantly enhanced by CIP and the technique is independent of the type of recording device. In the SPM imaging context, CIP can be understood as an a posteriori "sharpening of the effective experimental scanning probe tip" by computational means. It is now possible to remove multiple scanning probe mini-tip effects in images from 2D periodic arrays of physical objects that either self-assembled or were created artificially. Accepted within the scientific community is by now also the fact that SPM tips can change their shape and fine structure during the operation of a microscope and, thereby, obfuscate the recorded images in systematic ways. CIP restores much of the smeared out information in such images. The adaptation of a geometric Akaike Information Criterion from the robotics and computer vision community to the unambiguous detection of 2D translation symmetries enabled much of our recent progress. In the main body of this book chapter, we discuss this adaptation and briefly illustrate its utility on an example.

## 1. Introduction

Crystallographic Image Processing (CIP) is an averaging technique originally developed for the analysis of biological macromolecules imaged with transmission electron microscopes (TEMs) [1-3]. Key to this kind of averaging is the presence of multiple copies of identical objects with well defined two-dimensional (2D) point symmetries that are arrayed in a more or less regular 2D periodic manner within an image that was obtained by a noisy imaging process. Individual defects in the array, such as molecular or atomic vacancies, for example, do not present obstacles to the successful application of CIP when they are not too numerous. The CIP averaging technique can, however, be applied to experimental images from genuinely paracrystalline [4] arrays only as a zero order approximation.

It is typically the structure (including the shape, size and symmetry) of the microscopic object itself that is of interest to science. A 2D periodic array of this object then only serves the purpose of making the averaging within CIP feasible. As a result of the averaging, the signal to noise ratio of the image intensity is significantly enhanced so that conclusions on the object structure (shape and size) can be drawn with more confidence. These conclusions can in turn be related to the object's physical properties or biological functions. The symmetry of the objects in a CIP processed/symmetrized image is determined by the plane symmetry group that has been applied.

As far as increasing signal to noise ratios in experimental images from 2D periodic arrays is concerned, CIP is far superior to traditional Fourier filtering [5] because all of the plane symmetries that the array ideally possesses are taken into account. Crystallographic image processing for scanning probe microscopy (SPM) is conceptually equivalent to an a posteriori sharpening of the effective experimental scanning probe tip procedure – a processing step that is desirable for any kind of noisy SPM image from a 2D periodic array of objects. The point spread function of a SPM or TEM can be derived on the basis of CIP in a straightforward manner from images that were recorded from 2D periodic calibration samples [6-8]. The inverse of this point spread function can then be employed to correct subsequently recorded images [8]. In the case of scanning transmission electron microscopy, CIP allows for the removal of so called "electron probe tail artifacts" from atomic resolution images.

Portland State University's Nano-Crystallography Group pioneered the adaptation of CIP to the analysis of noisy 2D periodic images that were recorded with scanning probe microscopes [6-14]. The underlying object arrays were composed of either metal-organic or organic molecules on strongly or weakly interacting substrates [8-11] or the square mesas of a commercial SPM calibration sample [8]. While the former objects were imaged with scanning tunneling microscopes (STMs), the commercial calibration sample was imaged with an atomic force microscope (AFM) in order to demonstrate the influence of different recording parameters on this instrument's point spread function [8]. For practical purposes, we found that either square or circular selections of arrays in experimental images encompassing several tens to about one hundred repeating objects sufficed.

The mathematical foundations of CIP are comprehensively discussed in refs. [2,3]. For the convenience of the readers, we discussed some of the crystallographic foundations of CIP in refs. [6,7] and in the appendices to refs. [12,14]. Note that there are also two Master of Science theses on this subject [8,11] openly accessible (directly from within this book chapter), which feature good descriptions of the mathematical and crystallographic foundations as well.

In recent years, researchers in the SPM community began to appreciate that the recording tip of a SPM can change during the operation of the microscope [15,16] and, thereby, obfuscate details in the images in systematic ways. Significant image obfuscations were observed for example in a STM study and it was suspected that the tip may have picked up molecules from the 2D periodic array [15]. While an AFM tip was scanning across an inorganic crystal under water over several hours, atomic resolution details in the recorded images were suddenly lost. Subtle atomic rearrangements at the tip of the AFM were suspected to be the cause of that [16].

Our own recent developments [11-14] are particularly useful for correcting STM images for the imaging artifacts of double and multiple scanning probe mini-tips (or a blunt SPM tip in other words). These developments were enabled by our adaptation of a geometric Akaike Information Criterion (AIC) [17-23] to the unambiguous detection of the 2D Bravais lattice that an image of a 2D periodic array would most likely possess in the absence of experimental recording noise. As demonstrated with mathematical rigor in refs. [12,14], the recording of a 2D periodic image with multiple scanning probe mini-tips cannot change the translation symmetry (and 2D Bravais lattice) in the image. Note that both ref. [14] and the 2010 book chapter [6] mentioned in the abstract can be accessed openly from within this book chapter. Also note in passing that geometric AICs are known "to err a little bit on the side of caution" in comparison to both geometric Bayesian Information and geometric Minimum Description Length criteria, which could be employed as alternatives [21]. This means that geometric AICs have in real world applications a slight tendency to favor models with smaller deviation measures for a given set of "sophistication/generality" measures. Such models are in this book chapter the lower symmetric 2D Bravais lattices and their corresponding plane symmetry groups.

Combined with the traditional plane symmetry deviation quantifiers [3,6-8,11] of CIP, our geometric AIC is the key enabler for the correction of multiple scanning probe mini-tip effects in experimental SPM images. We will, therefore, in this book chapter concentrate on this criterion and illustrate its utility on an example. For the reader's convenience, we also provide a brief description of the traditional plane symmetry deviation quantifiers of CIP. A good college level textbook introduction to crystallography in 2D is ref. [24]. The reference text on this subject is ref. [25].

## 2. The three traditional plane symmetry deviation quantifiers of CIP

In order to determine the plane symmetry to which a 2D periodic image and with it the underlying more or less regular array of objects most likely belongs, one traditionally utilizes Fourier coefficient (FC) amplitude ($A_{res}$) and phase angle ($\phi_{res}$) residuals [3,6-8,11]. These residuals quantify how much an unprocessed image deviates from a symmetrized (CIP processed) one, and thus serve as figures of merit for determining which plane symmetry group best models both the experimental image and the underlying sample that has been imaged. There is, however, currently no fully objective way to use these two residuals (deviation quantifiers) to assign the correct plane symmetry group to a noisy image. This is because higher symmetric plane groups (such as *p4mm*) possess a higher multiplicity of the general position (per primitive unit cell) than their subgroups (such as *p4*, *c2mm*, and *p2mm*). These subgroups are formed from their supergroup by the removal of certain symmetry operations. Relations such as this are known as inclusion relations.

Whenever there are inclusion relations, one cannot simply base a decision on which model best describes the data in the absence of noise by a deviation quantifier alone [17]. This is because the weakest model, in our case the plane group with the lowest symmetry, will always have the smallest distance measure (regardless on how it is defined). Application of a geometric AIC [18-23] allows one to make decisions on the basis of information theory because the sophistication/generality of all of the models to be ranked enter the calculations in an appropriate way. In the absence of such a criterion, one would generally conclude that a particular plane symmetry group is the more likely group whenever the FC residuals of an image are not "significantly larger" for that group than for its respective subgroups. There is however no quantification of what significantly larger may mean in general or in any particular case.

In addition to the FC amplitude and phase angle residuals, it is also customary to utilize the so called $A_o/A_e$ ratio [3,6-8,11] for those six plane symmetry groups that possess systematic absences [25]. This ratio is defined as the average amplitude of the FCs that are forbidden in that plane symmetry group, $A_o$ (subscript $_o$ for *odd*), divided by the average amplitude of all other observed FCs, $A_e$ (subscript $_e$ for *even*). For plane symmetry groups *pg*, *cm*, *p2mg*, *p2gg*, *c2mm*, or *p4gm*, this ratio is helpful for informed plane symmetry model selections because it is zero for strict symmetry adherences. For noisy 2D periodic images, a larger value of this ratio makes it more unlikely that the corresponding group is the correct plane symmetry. Because of inclusion relations [17], the application of this plane symmetry deviation quantifier is also not fully objective. Note that the three traditional plane symmetry deviation quantifiers of CIP are all based on Fourier transforms so that decisions are made in reciprocal space.

## 3. Constraints for our geometric AIC

Within the robotics and computer vision communities, model selections involving non-disjoint classes of models often employ geometric AICs because these criteria deal with inclusion relations properly and their application supports unambiguous decisions in the presence of Gaussian noise of mean zero [18-23]. While the ultimate goal of applying a geometric AIC in conjunction with CIP would be to unambiguously identify plane symmetries in noisy 2D periodic images, we present here the basics of a geometric AIC procedure which unambiguously determines the underlying 2D translation symmetry in such images. These translation symmetries are represented by the five types of lattices that are commonly used by crystallographers and are known as the 2D Bravais lattices [25], see Fig. 1.

We adapted for our purposes a geometric AIC that was originally developed by Kenichi Kanatani and coworkers [22,23] for the classification of a computer mouse drawn quadrilateral as one of the quadrilaterals with at least the constraints of a trapezoid (trapezium outside of North America). In the Euclidean plane, the positions of the vertices for any such quadrilateral (i.e. trapezoid, parallelogram, rectangle, rhombus, and square) is subject to one or more of the constraints listed in Fig. 2. Within a set of constraints, the relations are algebraically independent.

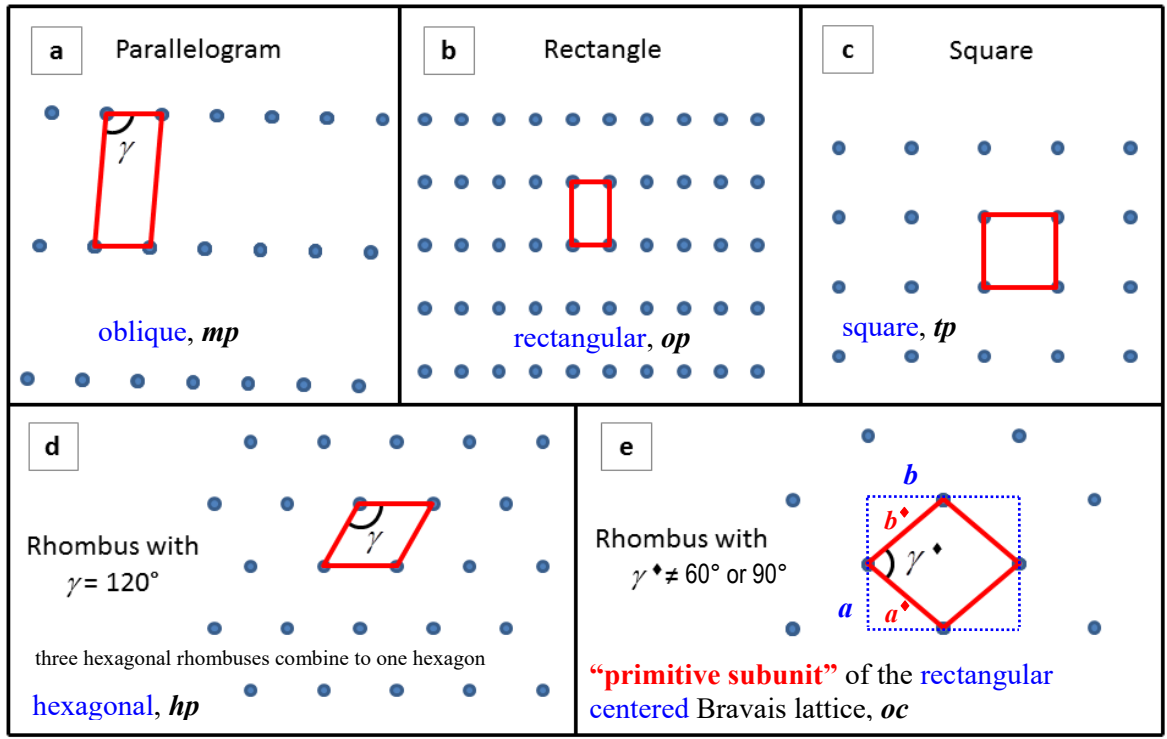

**Fig. 1 (a-d)** Unit cell shapes for the four primitive 2D Bravais lattices. **(e)** Shape of the "primitive subunit" (red) of the rectangular centered 2D Bravais lattice. The shapes of these unit and subunit cells correspond to the shapes of quadrilaterals with two or more geometric constraints on their vertices (Fig. 2). The points in (a-d) are lattice points. The crystallographic standard conventions [25] for the four primitive unit cells are that the x-axis points downwards and the y-axis points to the right. This results in a unit cell angle of $\gamma > 90°$ for the parallelogram in (a) and $\gamma = 120°$ for the hexagonal rhombus in (d). The angle $\gamma^\blacklozenge$ is by crystallographic convention neither 60° nor 90°. The common name of the quadrilateral is given in each panel. The crystallographic name of the corresponding Bravais lattice and its standard abbreviation are given at the bottom of each panel. As outlined in section 4, a few changes suffice to turn this set of panels into representations of 2D Bravais lattices in reciprocal (Fourier) space. Modified after a sketch in ref. [11].

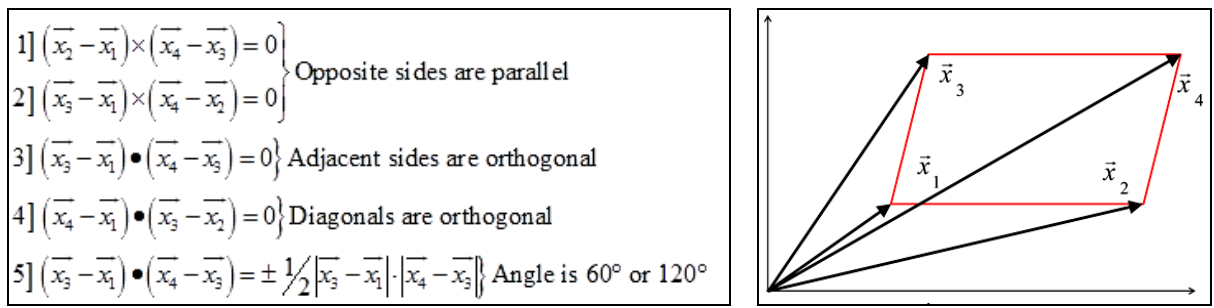

**Fig. 2** Geometric constraints on the vertices of quadrilaterals in the Euclidean plane. One pair of opposite sides is parallel to each other in a trapezoid. This constitutes a geometric constraint. The quadrilaterals that we are interested in (see Fig. 1) possess more than one constraint and inclusion relations with respect to their shapes (see Fig. 3a). Modified after sketches in ref. [11].

A square, for instance, is subject to constraints {1, 2, 3, 4} in Fig. 2. A rectangle is only subject to a subset of the square's constraints, i.e. {1, 2, 3}. This constitutes an inclusion relation [17]. We took four types of constraints for the quadrilaterals listed above, all involving cross products or dot products, from refs. [22,23] and used them to define the shapes of the unit cells of three of the four primitive 2D Bravais lattices. For the case of a special rhombus with "interior angle" of 60° (and lattice defining angle of 120° per crystallographic convention [25]), which is of importance to our particular application as it represents the hexagonal Bravais lattice, Fig. 1d, we came up with a fifth constraint as listed in Fig. 2. Our fifth constraint allows one to distinguish between rhombuses with and without an interior angle of 60° or 90°. The shape of the primitive "subunit" of the rectangular centered Bravais lattice [25] is represented by a general rhombus (with interior angel ($\gamma^\bullet$) other than 60° or 90°, Fig. 1e), for which the geometric shape constraints were already given in refs. [22,23]. Figure 3a represents the "geometric hierarchy" of quadrilaterals arranged by the total number of geometric constraints. The shapes of five types of quadrilaterals (in the upper part of Fig. 3a) correspond exactly to the five possible shapes of units and subunits of 2D periodic arrays, Fig. 1.

It is straightforward to construct an inclusion relation diagram that is equivalent to Fig. 3a for the symmetries of the four primitive unit cells and the one subunit cell of the 2D Bravais lattices in purely crystallographic terms, Fig. 3b. A plane symmetry group containing a holohedral 2D point symmetry is commonly referred to as the 'holohedral plane symmetry group' and is also the plane symmetry group of the corresponding 2D Bravais lattice [25]. For the inclusion relation diagram of these holohedral plane symmetry groups, the non-translational symmetry operations which are generators for the group, i.e. which combine with the group's translation vectors to generate all symmetry operations of the group, can be taken as symmetry constraints on the crystallographic unit and subunit cells.

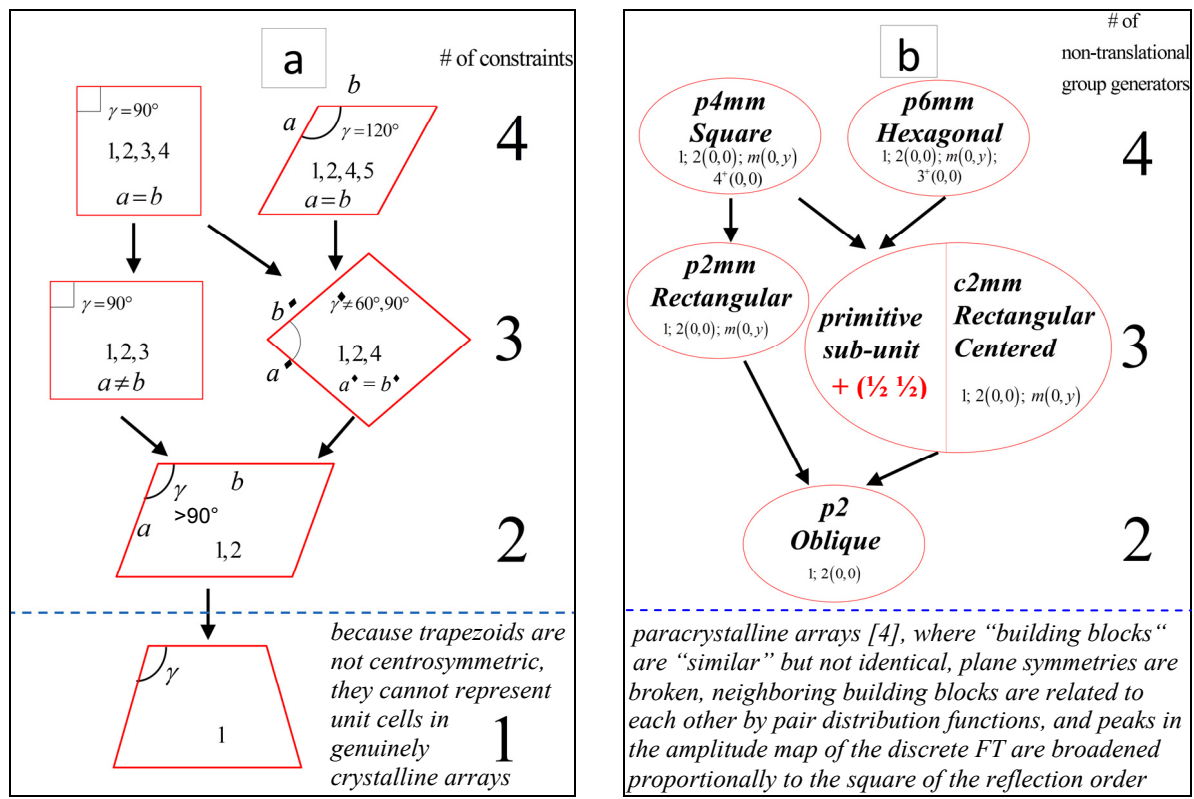

**Fig. 3 (a)** Inclusion relation diagram of the geometric hierarchy of quadrilaterals. The number and list of constraints (from Fig. 2) on the shape of the quadrilaterals are given for each level of the hierarchy. Connectors between quadrilaterals indicate that the constraints of the bottom quadrilateral are a subset of the constraints of the top quadrilateral. **(b)** Holohedral plane symmetry group inclusion relation diagram of the crystallographic hierarchy of the 2D Bravais lattices. Plane symmetry groups which contain a holohedral point group are listed along with the 2D Bravais lattice type. Sets of non-translational group generator operators (and their location within the standard crystallographic unit cell [25,26]) that highlight the inclusion relations are also given. Note that all unit or subunit cells in (b) are centrosymmetric, i.e. contain two-fold rotation points at position (0,0). While (a) refers to the shape of one unit or subunit cell individually, (b) refers to the shapes of all such cells collectively as two translations are involved. Entities below the dashed (blue) lines at the bottom of both panels are outside the scope of traditional CIP. Modified after sketches in ref. [11].

For the rectangular centered Bravais lattice with plane symmetry *c2mm*, the generating point symmetry operations at the provided locations in Fig. 3b combine with three translation, i.e. (1,0), (0,1), and (½,½), rather than the usual two for all primitive Bravais lattices. The centering has the effect of doubling the unit cell area in direct space, but does not constitute a crystallographic constraint by itself. (Endnote [26] continues the discussions of the rectangular centered unit cell.) The primitive subunit of a lattice with plane symmetry group *c2mm* possesses the shape/symmetry of a general rhombus with an internal angle ($\gamma^\bullet$) other than 60° or 90°, see Figs. 1e and 3a. A square and a hexagonal rhombus can be considered to be special kinds of rhombuses with the added constraint of an internal angle of 90° or 60° (lattice defining angle of 120° [25]), respectively. Restricting the internal angle to one of these two values allows one to "move up" in the hierarchy in Fig. 3a. Conversely, relaxing the internal angle constraint leads to a descent in this hierarchy.

The shapes of four of the quadrilaterals that are subject to up to four of the five geometric constraints on their vertices in Fig. 3a (as listed in Fig. 2) correspond directly to the shapes/symmetries of the unit cells of four of the five 2D Bravais lattices, Figs. 1a-d and 3b. For references to the fifth, the rectangular centered 2D Bravais lattice, Fig. 1e, we utilize the above mentioned primitive subunit [26]. The unit and subunit cell shapes/symmetries in which we are interested in can, thus, be defined in terms of either geometric constraints on the vertices of the corresponding quadrilaterals (Figs. 2 and 3a) or constraints due to the non-translational generators of the plane symmetry holohedries (Fig. 3b). What is important is that the number of constraints is identical in both types of hierarchy trees.

Note that it is not important for our application of a geometric AIC how, exactly, the constraints are formulated. Of importance is only that the models for the experimental data, which are ranked by the geometric AIC with respect of their likelihood of representing the translation symmetry of noisy image data, are not disjoint. In other words, these models need to possess a hierarchy of constraints with the more "sophisticated" (symmetric) model possessing one more constraint than the less symmetric model, while also possessing all of the constraints of the more general model.

## 4. Geometric AIC for the unambiguous detection of 2D Bravais lattices

To quantify deviations from translation symmetry models, we utilize a five member set of squared distance measures $J^i$ ($i = 1 \ldots 5$), one for each 2D Bravais lattice. This type of distance measure quantifies how much experimental image data differs from its model counterparts, i.e. from the five crystallographic translation symmetry models, Figs. 1 and 3b. In order to deal with inclusion relations within this set of models properly, we utilize a geometric AIC. Useful relations are derived from the equation for this geometrical AIC: (*i*) the ratios of *J* residuals that allow for the selection of a model with respect to the other models within the same hierarchy branch (Fig. 3), (*ii*) the "information content" of an image with respect of the process of selecting a geometric model, and (*iii*) the "confidence level" for the preference of a geometric model from a set of models with inclusion relations. Note that our unambiguous identification of translation symmetry in an experimental image from a 2D periodic array of objects results from both the application of a set of squared distance measures and their usage within the applicable geometric AIC.

It is irrelevant whether the geometric AIC is employed in direct or reciprocal (Fourier) space. For computational efficiency, we prefer to define our translation symmetry deviation quantifiers in reciprocal space. The shapes and symmetries (or types) of unit or subunit cells of Bravais lattices are, of course, unaffected by Fourier transforms between these two spaces. The major change required in Fig. 1 to represent 2D Bravais lattices in reciprocal (rather than direct) space would be to replace all references to 120° with 60° and vice versa. Statements such as $\gamma > 90°$ in Fig. 1 would need to be changed into $\gamma^* < 90°$. The lower-left or "most-left" vertex of each of the quadrilaterals in Fig. 1 could then represent the position of the (0,0) FC peak in the discrete Fourier transform (FT) amplitude map.

### 4.1 The five member set of translation symmetry deviation quantifiers

It is not only practical to determine both the positions for the vertices of the unit or subunit cells and the uncertainties in these positions in reciprocal space, but also highly advantageous. This is because both the positions of the (1,0), (0,1), and (1,1) FC peaks and their (peak) shapes in the FT amplitude map result from translation averaging over all of the selected unit or subunit cells in an experimental image. We have straightforward computational access in the amplitude map of a FT to the average values of the reciprocal unit ($a^*$, $b^*$ and $\gamma^*$) or subunit ($a^{\bullet *}$, $b^{\bullet *}$ and $\gamma^{\bullet *}$) cell parameters including their error bars. (An endnote mentions a crystallographic peculiarity of the two lattice points containing rectangular centered Bravais lattice in reciprocal space that is of no further consequence to our procedure [27].)

Five symmetrized unit or subunit cell shapes (and the corresponding lattice parameters) are obtained in reciprocal space (from the same experimental data) by CIP enforcing the holohedral plane symmetry group that belongs to each of the 2D Bravais lattices [28]. A plane symmetry group is enforced by averaging the symmetry related FCs accordingly. As a result, FC peak heights change in the FT amplitude map. New high order FC peaks may also appear. Another result of the symmetrization process is that peaks in the FT amplitude map change their positions [28]. These changes naturally affect the positions of the (1,0), (0,1), and (1,1) FC peaks in the FT amplitude map, resulting in the needed five member set of unit or subunit cell parameters in reciprocal space and, hence, a symmetrized unit or subunit cell shape set. The applied CIP procedures ensure that all unit or subunit cells have nearly the same area. On the basis of the derived reciprocal space unit or subunit cell parameters, we calculate a five member set of translation symmetry deviation quantifiers that needs to be considered as part of a geometric AIC for the decision on the underlying translation symmetry of a 2D periodic array for which we have an experimental image. As will become clear below, the absolute magnitude and physical unit of the two (reciprocal lattice) translation vectors per unit or subunit cell in the set is not important because only ratios of *J* values need to be considered within our geometric AIC procedure.

Trapezoids cannot represent unit or subunit cells of 2D Bravais lattices because they are not centrosymmetric. The trapezoid's shape (and its place in the geometric hierarchy of quadrilaterals, Fig. 3a), is, however, useful to us because it is obtained by extending or shortening one side of a parallelogram (or any of the other quadrilaterals that we are interested in) by shifting one vertex forward or backward along the line of its side, see Fig. 4a. We can use the trapezoid's shape to "sample the uncertainty" in the determination of the positions of the above mentioned (1,0), (0,1), and (1,1) FC peaks in the amplitude maps of the five symmetrized FTs that we obtained from the applied CIP procedures as mentioned above [28]. The necessary extra shifts of the vertices (which make trapezoids) out of the quadrilaterals (that represent both the average unit or subunit cells in reciprocal space and the primitive "translation units" of our five translation symmetry models) can be taken to match full widths at half maximum of the (1,0), (0,1), and (1,1) FC peaks in the set of symmetrized FT amplitude maps. This idea is illustrated in Fig. 4a on a parallelogram.

Variations in the shapes of these FC peaks are due to (direct space) irregularities in the 2D periodic array and image, see endnote [29], and influence the sampling of the positions of these peaks. The quadrilaterals that are defined by the (0,0), (1,0), (0,1), and (1,1) FC peaks in the FT amplitude map of a 2D periodic image are on theoretical grounds always centrosymmetric. A total of 16 trapezoids could be produced from a centrosymmetric quadrilateral by shifting vertices as described above (and shown in Fig. 4a) in order to facilitate the sampling of FC peak positions. The reciprocal space equivalents of 12 of them are currently utilized by us for the calculation of our five member set of translation symmetry deviation quantifiers $J$, see endnote [29] as well as appendix A of ref. [11] for the applicable equations.

Figure 4b illustrates the distances ($d_i$) from the vertices of experimental data, $x_i$, of an average unit or subunit cell in an FT amplitude map of an experimental 2D periodic image to the corresponding model vertices, $x_i'$, of one of the quadrilaterals that possess one of the shapes of the unit or subunit cells of the four higher symmetric 2D Bravais lattice models (Figs. 1b-e). The sum of the squares of these distances is our translation symmetry deviation quantifier:

$$J = |x_2 - x_2'|^2 + |x_3 - x_3'|^2 + |x_4 - x_4'|^2 \tag{1}$$

We use prime notation for our model unit or subunit cell shapes to differentiate them from the experimental unit cell shape, which is given without primes, Fig. 4b, and for generality taken to be a parallelogram. What is important for calculations of $J$ values is that the model unit or subunit cell area is nearly kept constant and matches that of the experimental unit cell area. (This is ensured by the CIP symmetrization procedure described above, which delivers a set of reciprocal lattice or primitive sublattice parameters with error bars [29].) The uncertainties of determining the position of three of the vertices of each model quadrilateral (corresponding to (1,0), (0,1) and (1,1) FC peak positions) are sampled as discussed above, but the sampling could be improved by the utilization of other schemes.

We currently calculate 12 preliminary $j$ values in an analogous manner to relation (1) for each final value of our translation symmetry deviation quantifier. A final $J$ value is obtained for each of the five models by averaging over the corresponding preliminary $j$ values. Error propagation calculations lead from the position uncertainties of three vertices of non-centrosymmetric quadrilaterals (i.e. trapezoids) in reciprocal space, via error bars on the reciprocal lattice or sublattice parameters, to error bars on each of the five translation symmetry deviation quantifiers $J$.

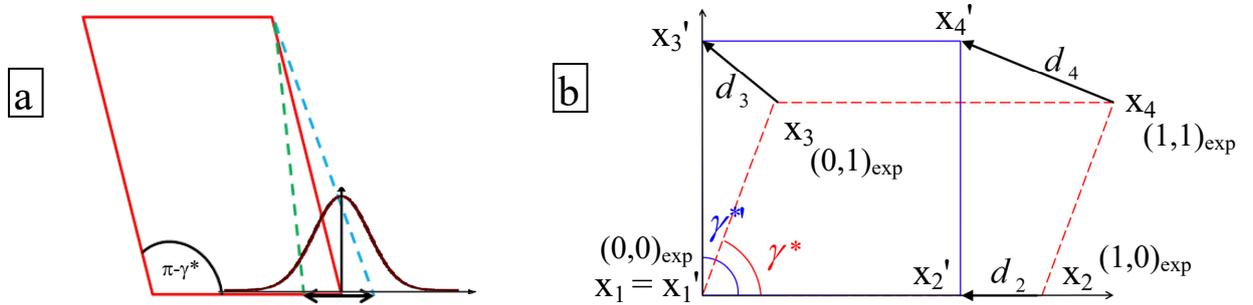

**Fig. 4 (a)** Parallelogram from which two trapezoids are created by shifting a vertex forwards or backwards parallel to one of its sides. These kinds of trapezoids are used for the calculation of preliminary $j$ values that contribute to a final $J$ value. The Gaussian distribution function inset at the shifted vertex illustrates that a trapezoid's shape samples the position of that vertex. **(b)** Illustration of our translation symmetry deviation quantifier $J$ (as obtained from the corresponding preliminary $j$ values). The sums of the squared differences between the positions of the (0,0), (1,0), (0,1), and (1,1) FCs in the amplitude map of the FT of an experimental image and the positions of their model counterparts in FT amplitude maps (of versions of that image) that were symmetrized to holohedral plane symmetry groups by CIP [28] are squared measures of the deviation of the experimental data from the five possible translation symmetries. For specificity, the vertex distances between an (experimental) parallelogram and a (model) square of the same area are shown, but the illustration of $J$ for the other four cases is analogous. Modified after similar sketches in ref. [11].

Since there are five 2D Bravais lattices (in both direct, Fig. 1, and reciprocal space) there is in our procedure a five member set of $J$ residuals that quantify deviations from the crystallographic translation symmetries. In other words, the set of $J$ residuals are squared distance measures for how much the shape of an average unit or subunit cell of a 2D periodic image differs from the shape of the corresponding quadrilaterals that represent the "primitive translation entity" of the corresponding 2D Bravais lattices. Only ratios of $J$ values will be of interest for the application of the geometric AIC below so that we normalize the set of translation symmetry deviation quantifiers by dividing all members of the set by the $J$ value for the parallelogram, which will typically be the smallest. When we assume that the relevant FC peak positions are (at least in a first order approximation) subject to independent Gaussian noise of mean zero and variance $\sigma^2$, it follows that the ratio $J/\sigma^2$ must approximately obey a $\chi^2$ distribution with $rN - n$ degrees of freedom [18,20].

### 4.2 The geometric AIC and parameters for our geometric model selection procedure

For comparing two non-disjoint models $S^m$ and $S^l$, with residuals $J^m$ and $J^l$ relative to the same experimental image data, we determine which model is the better fit by comparing their geometric AIC values (rather than their $J$ residuals alone). We do this (as already mentioned in section 2) because any set of residuals will favor the less constrained (more general/less sophisticated) model whenever there are inclusion relations [17]. ($S^m$ is a more sophisticated/symmetric model (superscript $^m$ for *more*) than $S^l$ (superscript $^l$ for *less*) and possesses the higher number of constraints $L^m$. $L^l$ is applicable to model $S^l$ so that $L^m > L^l$). The geometric AIC for a model $S$ is defined as follows [18,20]:

$$AIC(S) = J + 2(DN + n)\sigma^2 \qquad (2)$$

The data has $N$ degrees of freedom and is subject to independent Gaussian noise with standard deviation $\sigma$ and mean zero. The model $S$ is represented by a manifold of dimension $D$ with $n$ degrees of freedom.

The geometric AIC measures the predictive capacity of a geometric model in information theory terms [18-21]. If the image data that needs to be judged with the help of the geometric AIC is the positions of the vertices of an arbitrary quadrilateral (under no constraint) in the Euclidian plane, then each vertex has 2 degrees of freedom, giving the entire data a degree of freedom $N = 2$ times $4 = 8$, which is also the dimension of the data space. An arbitrary quadrilateral can, thus, be defined as a single point in an 8-dimensional data space. If a model $S$ for the data has additional constraints, it will define a sub-manifold of the data space with dimension $D$ and $n$ degrees of freedom (i.e. the degrees of freedom of a vector which parameterizes the $L$ constraints).

For example, a square model can be defined as a point, constrained to lie on the surface of a manifold, $S$, which is a linear subspace of the 8-dimensional data space. The co-dimension of a linear subspace, $r$, is defined as follows: co-dimension (subspace) = dimension (data space) – dimension (subspace). In particular, the sub-manifold representing the model space for squares possesses dimension $D = 8 - 4 = 4$ (from $D = N - L$ where $L$ is the number of constraints which can be gleaned from Fig. 3), degree of freedom $n = 4$ ($n = L$), and co-dimension $r = 8 - 4 = 4$ ($r = N - D$). Similarly, a rectangular model possesses $D = 5$, $n = 3$, and $r = 3$. Table 1 lists the parameters of the applicable AIC for the five quadrilateral shapes that we are interested in as models for translation symmetric unit or subunit cell shapes.

**Table 1** Geometric parameters that enter into the calculation of the applicable geometric AIC for our unambiguous translation symmetry identification procedure. The degree of freedom (or dimension) of the data space ($N$) is eight.

| | model dimension (D) | degree of freedom (n) | co-dimension (r) | # of constraints (L) |
|---|---|---|---|---|
| parallelogram / p2 / mp | 6 | 2 | 2 | 2 |
| rectangle / p2mm / op | 5 | 3 | 3 | 3 |
| general rhombus / c2mm / subunit of oc | 5 | 3 | 3 | 3 |
| square rhombus or rectangle / p4mm / tp | 4 | 4 | 4 | 4 |
| hexagonal rhombus / p6mm / hp | 4 | 4 | 4 | 4 |

### 4.3 Ratios of J residuals that allow for the selection of one non-disjoint model over another

When $S^m$ is a more sophisticated model than $S^l$ (as obtained by imposing an additional symmetry/geometric constraint on model $S^l$), it is shown in refs. [18,20] and backed up by experimental/simulated evidence [19,21] that, in the first order, the following is an unbiased estimator of the squared noise level for both models:

$$\sigma^2 = \frac{J^l}{r^l N - n^l} \qquad (3)$$

where $r$ is the co-dimension of the model and the relation is valid as long as $rN - n$ is larger than zero.

As the squared noise level is the same for both (the $^l$ and the $^m$) models, we do not need to estimate it, but it needs to be Gaussian distributed noise of mean zero. For other types of noise, relations (2) and (3) may either be fulfilled approximately or the whole geometric AIC approach must be generalized [18,20].

A more constrained model $S^m$ will be favored over a less sophisticated model $S^l$ if it has a smaller geometric AIC,

$$AIC(S^m) < AIC(S^l) \qquad (4a)$$

For

$$AIC(S^m) > AIC(S^l) \qquad (4b)$$

the less constrained model will be favored over the more sophisticated model. No decision on which model is favored is possible for

$$AIC(S^m) = AIC(S^l) \qquad (4c)$$

Combining equations (2), (3) and (4a), we are left with the following relation:

$$\frac{J^m}{J^l} < 1 + \frac{2(D^l - D^m)N + 2(n^l - n^m)}{r^l N - n^l} \qquad (5)$$

which must be true for the more sophisticated model $S^m$ to be favored over the less symmetric model $S^l$ [18,20].

For the five quadrilaterals in the upper part of Fig. 3a, equation (5) reduces with variable substitutions according to section 4.2 (as summarized in Table 1) to the following inequality:

$$\frac{J^m}{J^l} < \frac{2L^m - L^l}{L^l} \quad (6a)$$

where $L^m$ is the number of constraints on the more symmetric (sophisticated/constrained) model (superscript $^m$) and $L^l$ is the number of constraints on the lesser (i.e. more general) model (superscript $^l$). For comparing the shapes of two quadrilateral models with an inclusion relation with respect to the same image data, the more constrained model will be favored if relation (6a) is satisfied [22,23]. The less constrained model will be favored otherwise, i.e. when

$$\frac{J^m}{J^l} > \frac{2L^m - L^l}{L^l} \quad (6b)$$

### 4.4 Information content of an image and confidence level of a geometric AIC based preference

On the basis of the geometric AIC for two quadrilateral models with an inclusion relation, the information content, $K$, of an experimental image with respect to the process of model selection was introduced in refs. [19,20] as:

$$K^m = \sqrt{\frac{AIC(S^m)}{AIC(S^l)}} = \sqrt{\frac{r^l N - n^l}{(2D^l + r^l)N + n^l}\left(\frac{J^m}{J^l} + \frac{2(D^m N + n^m)}{r^l N - n^l}\right)} \quad (7)$$

Whenever the inequality (4a) is fulfilled, $K^m$ is smaller than unity and the more sophisticated model represents the image data better than the less constrained model.

In the special case of $K^m = 1$, no decision can be made if the more sophisticated model or the less symmetric model is the better representation of the image data. This case corresponds to

$$\frac{J^m}{J^l} = \frac{2L^m - L^l}{L^l} \quad (6c)$$

and the confidence level, $C$, of a decision in favor of either model is zero percent.

The maximal (100 %) confidence level $C_{max}^m$ for a decision that the more symmetric model represents the image data better than the less constrained model is given by the condition $J^m = J^l$, which results in the relation:

$$K_{cri}^m = \sqrt{\frac{AIC(S^m)}{AIC(S^l)}} = \sqrt{\frac{N(2D^m + r^l) + 2n^m - n^l}{N(2D^l + r^l) + n^l}} \quad (8)$$

for the critical information content from which onward such a decision is possible. This entity can be utilized for the normalization of the maximal confidence level to certainty (100 %), yielding:

$$C^m = \frac{(1 - K^m)}{(1 - K_{cri}^m)} \cdot 100\% \quad (9)$$

Equation (9) is just an ad-hoc definition, whereby confidence levels of less than 50 % do not imply that relation (6a) and conclusions that follow form it are no longer valid. Our confidence level relation is intended to support the mental assessment of the $J^m/J^l$ ratio region within which relation (6a) is indeed valid, i.e. between unity and the numerical value on its right hand side. Low confidence levels mean according to relation (9) that one is close to (but still below) the maximal $J^m/J^l$ ratio where the more symmetric model is still favored over the lesser model. At low confidence levels, one needs to be extra cautious if the assumptions behind the application of the geometric AIC are really justified. Possible applications of the geometric AIC-ratio based information content concept are outlined below.

### 4.5 Towards automated procedures that obtain the best possible signal to noise ratio in 2D periodic images

As pointed out in ref. [20], relation (7) can serve as a "continuous measure" on the preferability of one model over another within some dynamical process. For the recording of images from 2D periodic arrays with known translation symmetry in the low electron dose mode in a dynamical transmission electron microscope (DTEM), for example, one could increase the electron dose gradually, in an incremental manner, while testing the recorded preliminary data using relations (7) to (9). In this way, one could determine whether or not increased dosing shifts these measures toward or away from the correct (a priori known) conclusion. The signal to noise ratio in these data would be expected initially to improve continuously with increasing electron dose so that the confidence levels of relations (9) will tend towards a maximal obtainable value. Increasing the electron dose further in increments after this maximal value is obtained does not make sense and could be an indication that the sample is now starting to be damaged structurally by dissipation of the energy that the electron beam deposited. This energy dissipation would result in continuously worsening signal to noise ratios as the electron dose is further increased. Similar effects would be obtained if the sample were drifting while the electron dose is increased incrementally. Again one should stop increasing the electron dose when the confidence level for the decision in favor of the (a priori) known translation symmetry begins to decline.

In terms of an application in robotics and computer vision [20], a robot could actively control a camera so that the information content of continuously recorded images tends to its maximal value in dependence of external influences such as changing scenery or lighting conditions. On a DTEM, a computer with "electron vision" could decide if further increasing the electron dose is likely to make the signal to noise ratio better or worse. When the robot on the DTEM tries to maximize the information content of the so far accumulated image data with respect to a geometric AIC-ratio based decision in a translation symmetry model selection process by using relation (9), it also tries to find the maximal signal to noise ratio for the (final) recorded image on the basis of the (a priori) known 2D Bravais lattice of the sample.

While this "vision" of future developments is in line with emerging trends in materials informatics [30], future DTEM applications will probably require generalized geometric AICs that can deal with Poisson noise. Such applications could then also be supported by a future geometric AIC for plane symmetry model selection.

## 5. Guidelines for the application of our geometric AIC

Taking up our example from section 3, we see from Fig. 3 that a square's shape is subject to four geometric (or crystallographic) constraints, while there are only three such constraints on a rectangle's shape. Then according to relation (6a), the square lattice is to be favored on information theoretical grounds over the rectangular lattice as translation symmetry for an experimental image as long as the residual $J^m$ for the square model is smaller than $\frac{5}{3} J^l$, the residual for the rectangular model. So a square would be the correct choice for the unit cell shape despite its typically larger residual as long as relation (6a) is fulfilled and there are only negligibly small systematic errors in the experimental data to which the geometric AIC methodology is applied.

To deal with the exclusion of the other three possible unit cell shapes, relations (6a) and (6b) need to be employed repeatedly. Confidence levels are to be calculated with relations (9). For example, in order to conclude that the shape of a unit cell is indeed a square, it must be preferred per geometric AIC over both a rectangle's and a general rhombus' shape, which in turn have to be both preferred over a parallelogram's shape. The general rhombus, on the other hand, must also be preferred over a hexagonal rhombus in this case.

It needs to be emphasized that there is no need for a rule of thumb or any arbitrarily introduced (subjective) threshold to compare the translation symmetry deviation quantifiers for two translation symmetry models in order to find out which one of them is more likely to present the data in the (extrapolated) absence of imaging induced noise. We have instead fully objective additive terms within the geometric AICs that account for the models' sophistication/generality. Any kind of threshold that is external to the problem is, therefore, not needed. We are justified to choose a square model or a rectangular model for the translation symmetry depending on whether the inequality relations (6a,b) are satisfied or not, respectively. In addition, we even have with relation (9) a robust information theory based measure for the confidence of our selection in favor of the more symmetric model.

Plugging the characteristic values for the square shaped unit cell and the rectangular shaped unit cell from Table 1 into relations (6a), (7), (8), and (9), and assuming that the ratio of their two J residuals is $\frac{4}{3}$, i.e. slightly higher than half way in the $J^m$ to $J^l$ ratio interval that allows for decisions in favor of the square unit cell, we obtain a confidence level of approximately 49 %. Note that this value being smaller than 50 % does by no means imply that the rectangle should be chosen over the square as unit cell model of the experimental data! Relation (6a) clearly mandates the opposite!

A low confidence level of about 15 % and smaller means that one should be cautious in preferring the square over the rectangle as average unit cell shape and perhaps reanalyze if relevant systematic errors (that need to be corrected as far as this is possible before the application of the geometric AIC) are really negligible in comparison to random Gaussian errors of mean zero that are dealt with effectively by the geometric AIC. With the hope of boosting the confidence level, one could try to devise a better FC peak position sampling scheme that does not involve trapezoids and make the translation symmetry deviation quantifiers more accurate and precise by that route [29]. For high confidence levels, on the other hand, one does not need to be too concerned about the accuracy and precision of the $J^m/J^l$ ratios.

We will present two examples of the unambiguous identification of square Bravais lattices in section 6 below. The example image that we will show explicitly involves simulated data in the context of the removal of a blunt SPM tip artifact. Due to that data having been simulated, confidence levels of geometric AIC based decisions on the underlying translation symmetry model will be essentially 100 %. The other example deals with experimental image data that has already been analyzed in ref. [11], so that there is no need to show the corresponding image here.

## 6. Removal of multiple mini-tip imaging artifacts in STM images

Although only a simulated example, Fig. 5 suffices for our purpose of showing how our unambiguous translation symmetry detection procedure enhances traditional CIP. The *p4*-symmetry in Fig. 5a is "symmetrically perfect" because we imposed this symmetry on an experimental "nearly-*p4*" STM image [31] using the CIP program CRISP [32].

In Figure 5b we have artificially constructed an image akin to what one would "see" with three STM mini-tips shifted laterally with respect to each other, constituting a blunt tip, and simultaneously scanning the same "sample" surface. Obfuscated images such as Fig. 5b have been discarded (or misinterpreted[e]) in the past, but CIP presents an alternative to recover the correct motif information from them. Our geometric AIC procedure can be applied to such images and CIP can be used for the purpose of their subsequent crystallographic averaging which results in the sharpening of the effective experimental scanning probe tip and restoration of the key features of the 2D periodic motif, Fig. 5c.

We find from visual inspection that the un-obfuscated "sample/image" in Fig. 5a, and the obfuscated image in Fig. 5b (as well as the CIP processed image, Fig. 5c, of course), possess the same translation symmetry, which is that of the square 2D Bravais lattice. While the obfuscation due to multiple SPM mini-tips cannot affect the translation symmetry, it may modify the point symmetry of the 2D periodic motif in the image significantly. This means that while the "sub-images" of the individual 2D periodic array objects may become smeared out beyond recognition, they will still be arranged in the very same 2D periodic manner throughout the whole image. These facts are illustrated by Fig. 5b. Tables 2a and 2b list the traditional plane symmetry deviation quantifiers of CIP from the application of the CRISP program [32] to this figure.

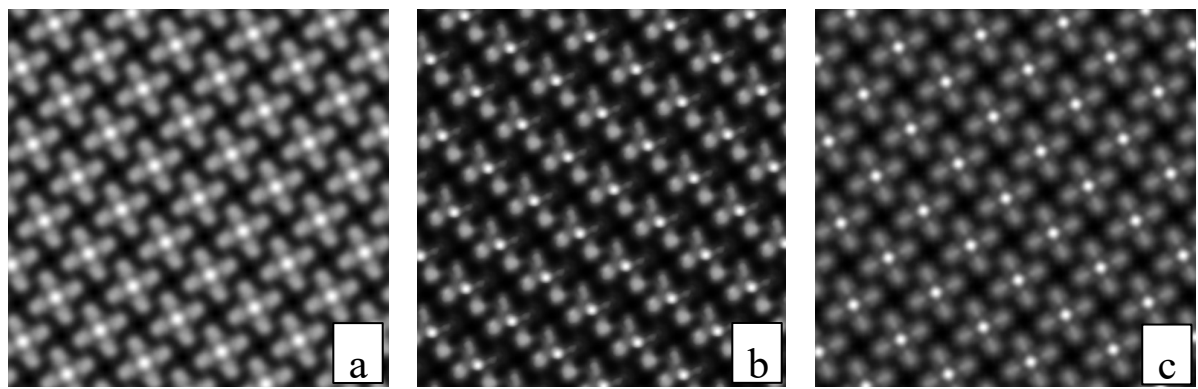

**Fig. 5 (a)** A 512 by 512 pixel "simulated sample/image" whose *p4*-symmetry is known by design (on the basis of an experimental STM image in ref. [31]). The area of this image is approximately 80 nm$^2$ and contains a sufficient number of 2D periodic objects for CIP applications to be meaningful. **(b)** Simulation to model what three non-collinear mini-tips (or a blunt STM tip in other words) would have recorded from this sample, constructed in Photoshop. **(c)** CIP reconstruction of (b) with plane symmetry *p4* enforced.

| (a) | p2 | p1m1 | p11m | p1g1 | p11g | p2mm | p2mg | p2gm | p2gg | c1m1 | c11m | c2mm | p4 | p4mm | p4gm |
|---|---|---|---|---|---|---|---|---|---|---|---|---|---|---|---|
| $A_{res}$ | n.h. | 23.9 | 23.9 | 28.3 | 25.9 | 23.9 | 25.9 | 28.3 | 31.0 | 24.9 | 24.9 | 24.9 | **30.4** | 38.8 | 37.9 |
| $\phi_{res}$ | 23.3 | 20.3 | 21.3 | 26.0 | 24.8 | 35.5 | 41.8 | 38.0 | 32.1 | 17.5 | 12.8 | 22.6 | **24.2** | 34.8 | 32.7 |
| $A_o/A_e$ | n.h. | n.h. | n.h. | 1.5 | 0.8 | n.h. | 0.8 | 1.5 | 1.2 | 1.3 | 1.3 | 1.3 | **n.h.** | n.h. | 1.2 |

| (b) | p3 | p3m1 | p31m | p6 | p6mm |
|---|---|---|---|---|---|
| $A_{res}$ | 64.9 | 65.2 | 65.2 | 64.9 | 65.2 |
| $\phi_{res}$ | 27.3 | 35.0 | 33.6 | 34.4 | 40.4 |
| $A_o/A_e$ | n.h. | n.h. | n.h. | n.h. | n.h. |

**Table 2.** Traditional plane symmetry deviation quantifiers from the application of CRISP to Fig. 5b. The traditional plane symmetry deviation quantifiers for group *p4* are given in bold face. The letters n.h. stand for *not helpful* (because of being 0 always) **(a)** Non-hexagonal plane symmetry groups and **(b)** hexagonal plane symmetry groups.

Due to our sample/image (Fig. 5a) having been simulated assuming both ideal imaging conditions and a perfectly regular square array of molecules, the normalized *J* values of Fig. 5b are essentially unity for all of the non-hexagonal 2D Bravais lattices. Inequality (6a) is, thus, fulfilled for all of the non-hexagonal Bravais lattices, where superscript $^m$ stands for the square lattice in the rounds of decisions that results in its preference over both the rectangular (primitive) and the rectangular centered lattice (which followed the rounds of decisions of the preference of both models over the oblique model). This corresponds according to relation (9) to a confidence level of essentially 100 % for the preference of the square lattice as model for the translation symmetry in Fig. 5b. The normalized *J* value for the hexagonal model with respect to the "pseudo-experimental" data in Fig. 5b is, on the other hand, as large as 104.5, i.e. more than two orders of magnitude larger than its counterpart for the rectangular centered model. Inequality (6b) is obviously fulfilled (as $104.5 \gg {}^5/_3$), so that there is no doubt at all that a general rhombus is preferred over a hexagonal rhombus.

These kinds of (qualitative and quantitative) results are, of course, to be expected because the *J* residuals are, by definition, not affected by blunt scanning-probe tip effects. (We also know, of course, from the design history of Fig. 5b that it must possess the translation symmetry of the square Bravais lattice.) The assessment of the three traditional plane symmetry deviation quantifiers of CIP is with our detection of the correct translation symmetry simplified to a decision between the three plane symmetry groups that are based on a square lattice, i.e. *p4*, *p4mm*, and *p4gm*. Evaluating these quantifiers in Table 2a for these three plane symmetry groups, it is clear that the underlying plane symmetry is *p4*, which we, of course, also know to be true from the design history of Fig. 5b.

A typical result for an experimentally obtained (rather than simulated) image from an array of objects with an approximate (rather than a made perfect by symmetrization) square 2D Bravais lattice is that the normalized *J* values for the non-hexagonal translation symmetries increase slowly with the number of constraints on the translation symmetry models, while the corresponding value for the hexagonal model is about two orders of magnitude larger. The analysis of the experimental STM image from ref. [31] that we utilized above to create Figs. 5a,b resulted, for example, in the following normalized residuals: $J_{mp} = 1.00$, $J_{op} = 1.01$, $J_{oc} = 1.10$, $J_{tp} = 1.15$, and $J_{hp} = 105.3$ [11,29]. We use here the crystallographic standard [25] two-letter abbreviations of the 5 Bravais lattices, see Fig. 1, as subscripts. The confidence levels for decisions in favor of the square model over both the rectangular (primitive) model and the rectangular centered model are approximately 77 % and 93 %, respectively. The rectangular (primitive) model is favored over the oblique model with a confidence level of approximately 99 %. The confidence level for the decision in favor of the rectangular centered model over the oblique model is approximately 90 %.

It is, therefore, *quite* clear that the noisy experimental image from ref. [31] (that we utilized above to create Figs. 5a,b) and with it the underlying molecular array possesses a square Bravais lattice. This might, however, be a translational pseudo-symmetry [33]. Future developments and applications of geometric AICs for Laue classes and plane symmetry groups will enable us to revisit this assessment in an objective and comprehensive way.

## 7. Summary and Outlook

A detailed account of our adaptation of a geometric AIC to the tasks of the identification of the underlying translation symmetry of experimental SPM images of more or less regular 2D periodic arrays of identical objects was given. Our set of $J$ residuals, based on the positions of a few peaks in Fourier transform amplitude maps, is useful as a squared distance measure of crystallographic translation symmetries to be compared using a geometric AIC. This is because unlike the three traditional CIP figures of merit, our $J$ residuals are not affected by severe distortions of the periodic image motif by recordings with multiple mini-tips (also known as blunt tips).

The application of our geometric AIC makes the identification of the translation symmetry unambiguous, provided that relevant systematic errors are negligible and random errors possess at least approximately a Gaussian distribution with mean zero. Identifying the underlying translation symmetry in 2D periodic images improves the symmetry identification step in CIP analyses by reducing the possible plane symmetry groups to only those compatible with the 2D Bravais lattice identified. Decisions as to which plane symmetry most likely underlies noisy experimental image data from more or less regular 2D periodic arrays of self assembled objects are, however, currently not completely objective because one has to rely there on the three traditional plane symmetry deviation quantifiers of CIP.

Our future work will involve developing a geometric AIC procedure which can be used to identify the (full) plane symmetry group of a noisy experimental image directly (rather than being limited to its translation symmetry part).

**Acknowledgements**   This work was supported by both Portland State University's (PSU's) Venture Development Fund and the Faculty Enhancement program. A grant from PSU's Internationalization Council is also acknowledged. Collaborative work with former members of the author's/PSU's Nano-Crystallography Group, i.e. his MSc students Bill Moon and Taylor Bilyeu, as well as with his colleague (in PSU's Department of Physics) Prof. Jack Straton is gratefully acknowledged. Taylor Bilyeu is also thanked for creating the sketches that are shown here in modified form as Figs. 1 to 4. Professor Kerry Hipps of Washington State University at Pullman is thanked for the raw STM image that we used to create Fig. 5a. Professor Jack Straton is thanked for creating Fig. 5b. Paul DeStefano and Andrew Dempsey of PSU's Nano-Crystallography Group (in 2017) are thanked for critical proof readings of the manuscript. Professor emeritus Kenichi Kanatani of Okayama University is thanked for useful comments.